\documentclass{article}
\usepackage{a4wide}
\usepackage{amsmath,amsthm,amssymb,amsfonts}
\usepackage[T1]{fontenc}
\usepackage{mathptmx}
\usepackage[pdftex,pdftitle={Neutrino dimuon production and the dynamical determination of strange parton distributions}, pdfauthor={P. Jimenez--Delgado}, colorlinks=true, linkcolor=black, citecolor=black, urlcolor=black]{hyperref}
\usepackage[all]{hypcap}
\usepackage{graphics,graphicx}
\usepackage{rotating,subfigure}
\usepackage{float}
\usepackage[numbers,sort&compress]{natbib}
\usepackage[small]{caption}
\begin{document}
\thispagestyle{empty}
\begin{center}
\begin{minipage}{\textwidth}
\begin{flushright}
DO-TH 10/03\\
ZU-TH 03/10\\
\end{flushright}
\vspace{10mm}
\begin{center}
{\huge Neutrino dimuon production and the dynamical\\[0.3em] determination of strange parton distributions}\\[10mm]
{\large P. Jimenez-Delgado}\\[5mm]
\textit{Institut f\"ur Physik, Technische Universit\"at Dortmund, D-44221 Dortmund, Germany}\\[0.5em]
\textit{Institut f\"ur Theoretische Physik, Universit\"at Z\"urich, CH-8057 Z\"urich, Switzerland}
\end{center}
\begin{center}
{\small January 2010}
\end{center}
\vspace{3mm}\hspace{13.4mm}
\parbox{0.8\textwidth}{\small Utilizing recent neutrino dimuon production measurement from NuTeV the assumptions on the determination of the strangeness content of the nucleon within the dynamical approach to parton distributions are investigated. The data are found to be in good agreement with the predictions derived from our (GJR08) dynamical parton distributions, which have been generated entirely radiatively starting from vanishing strange input distributions at an optimally chosen low resolution scale. Further, the data induce an asymmetry in the strange sea which is found to be small and positive in agreement with previous results.}
\end{minipage}
\end{center}
\vspace{5mm}
Parton distributions and their implications have been recently studied within the dynamical parton model approach \cite{Gluck:2007ck, JimenezDelgado:2008hf}. Since the dynamical parton distributions at $Q^2 \gtrsim 1$ GeV$^2$ are QCD radiatively generated from {\em valencelike}\footnote{Valencelike refers to $a_f\!>\!0$ for {\em all} input distributions $xf(x,Q_0^2)\propto x^{a_f}(1-x)^{b_f}$, i.e., not only the valence but also the sea and gluon input densities vanish at small $x$.} positive definite input distributions at an optimally determined low input scale $Q_0^2\!<\!1$ GeV$^2$, the predicted steep small Bjorken-$x$ behavior of structure functions is mainly due to QCD--dynamics at $x \lesssim 10^{-2}$.  Alternatively, in the common ``standard'' approach the input scale is fixed at some arbitrarily chosen $Q^2_0>1$ GeV$^2$ and the corresponding input distributions are less restricted, for example, the observed {\em steep} small-$x$ behavior of structure functions and consequently of the gluon and sea distributions has to be {\em fitted} here. Furthermore the associated uncertainties encountered in the determination of the parton distributions turn out to be larger, particularly in the small-$x$ region, than in the more restricted dynamical radiative approach where, moreover, the evolution distance (starting at $Q_0^2<1$ GeV$^2$) is sizably larger (see \cite{Gluck:2007ck, JimenezDelgado:2008hf} and references therein).

As in previous dynamical determinations \cite{Gluck:1994uf, Gluck:1998xa}, since the data sets used are insensitive to the specific choice of the strange quark distributions, the strange densities of the \emph{dynamical} distributions in \cite{Gluck:2007ck, JimenezDelgado:2008hf} have been generated entirely radiatively starting from vanishing strange input distributions:
\begin{equation}
s(x,Q_0^2)=\bar{s}(x,Q_0^2)=0 
\label{nullstrange}
\end{equation}
at the \emph{low} input scale\footnote{In the ``standard'' case, where $Q^2_0\!>\!1$ GeV$^2$, the strange input distributions were chosen $s(x,Q_0^2)=\bar{s}(x,Q_0^2)= \tfrac{1}{4}\large(\bar{u}(x,Q_0^2)+\bar{d}(x,Q_0^2)\large)$, as is conventional \cite{Gluck:2007ck, JimenezDelgado:2008hf}.}. In order to investigate the plausibility of these \emph{assumptions}, we confront here predictions derived from dynamical distributions determined in this way with data which are particularly sensitive to the strangeness content of the nucleon. 

For this purpose we have chosen the latest and most precise measurements of neutrino dimuon production from $\nu_\mu$- and $\bar{\nu}_\mu$-iron deep inelastic scattering (DIS) interactions of NuTeV \cite{Mason:2007zz}. The measured cross--section requires the muon from the semileptonic charm decay to have an energy greater than 5 GeV, therefore the theoretical predictions need to be corrected from detector acceptance; the ``forward'' cross-section is given by \cite{Mason:2007zz}:
\begin{equation}
\frac{d\sigma^+}{dxdy}\!\left(x,y,E_{\nu(\bar{\nu})}\right)=\frac{G_F^2 M E_{\nu(\bar{\nu})}}{\pi} \;B_c \; {\cal A}\!\left(x,y,E_{\nu(\bar{\nu})}\right) \; \frac{d\sigma^{\nu(\bar{\nu})}}{dxdy}\!\left(x,y,E_{\nu(\bar{\nu})}\right),
\label{dimuon}
\end{equation}
where ${\cal A}$ denotes the acceptance corrections, and $\tfrac{d\sigma^{\nu(\bar{\nu})}}{dxdy}$ is the inclusive DIS cross-section for charged current charm production; fragmentation effects are included in the acceptance corrections as have been calculated in \cite{Kretzer:2001tc, Mason:2007zz} at NLO of QCD. The cross--section for (anti)neutrino charm production has been given up to NLO in \cite{Gottschalk:1980rv, Gluck:1996ve} within the ``fixed flavor number scheme'' (FFNS), i.e., besides the gluon, only the light quark flavors ($u, d, s$) should be included as (massless) partons within the nucleon (for a recent discussion on the treatment of heavy--quark masses see, for instance, \cite{Gluck:2008gs}); for consistency we will therefore use our (GJR08) NLO FFNS dynamical parton distributions of \cite{Gluck:2007ck} for the present investigations \footnote{The use of the expressions in \cite{Gluck:1996ve} together with ``variable flavor number scheme'' distributions, e.g. in \cite{Mason:2007zz}, is inconsistent.}. 

Since the NuTeV collaboration used an iron target, the mass $M$ in Eq.~(\ref{dimuon}) denotes the average nucleon mass and, moreover, the parton distributions have to be corrected for nuclear effects; we use the NLO corrections of \cite{deFlorian:2003qf}. These nuclear corrections were obtained using the previous set of NLO FFNS dynamical parton distributions of GRV98 \cite{Gluck:1998xa} (which are very similar to our GJR08, see \cite{Gluck:2007ck}) and therefore are especially suited for combination with our GJR08 NLO FFNS. As is clear from Eq.~(\ref{dimuon}), the absolute normalization of the measurements is directly related to the semileptonic branching ratio $B_c$, hence we will allow its value to float within the experimental error of the value used by the experiment $B_c\!=\! 0.099 \pm 0.012$ \cite{Mason:2007zz}, as is common for the (fully correlated) normalization errors in global QCD analyses. However, all our distributions prefer the highest allowed value.

\begin{figure}
\centering
\includegraphics[width=\textwidth]{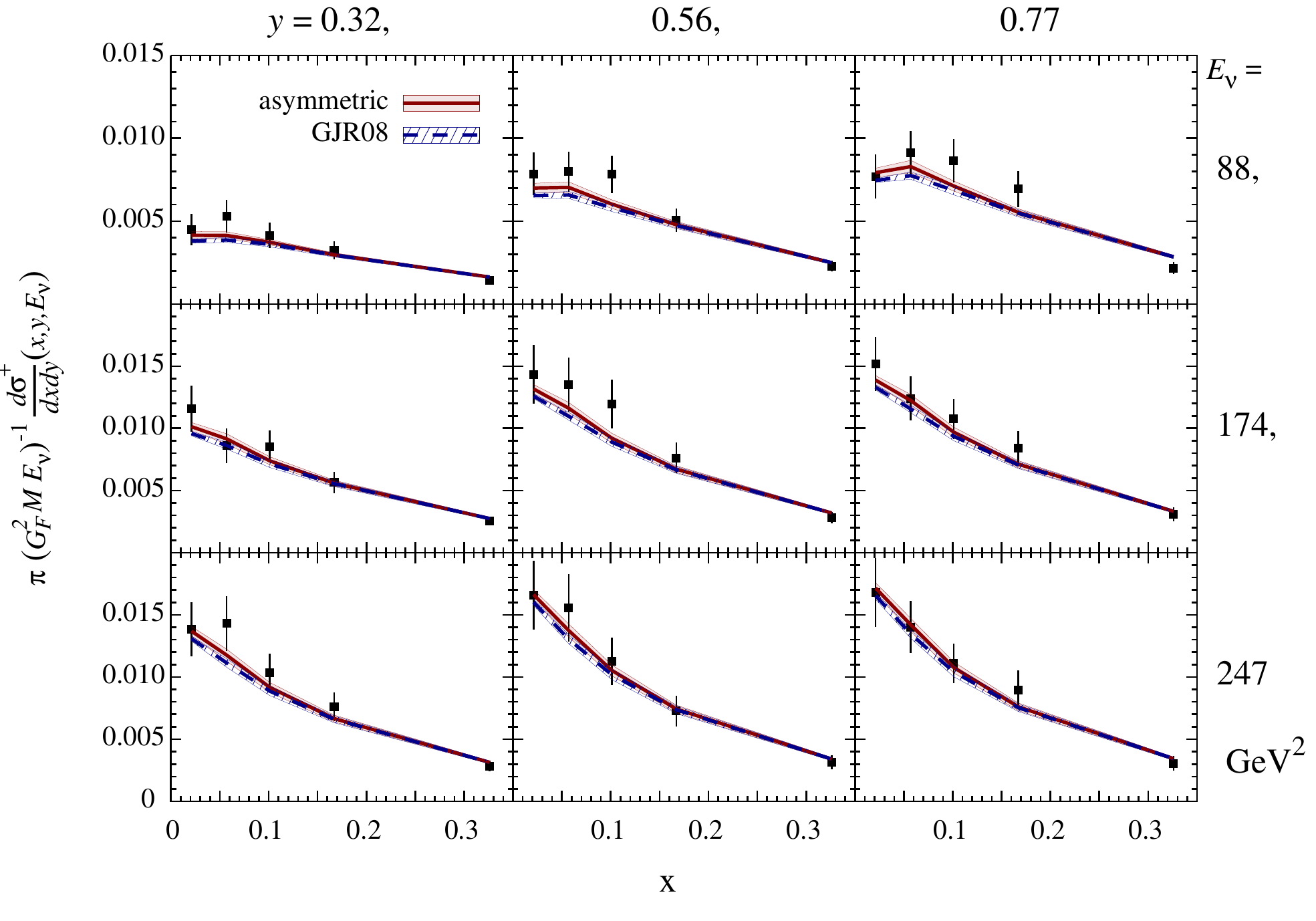}
\includegraphics[width=\textwidth]{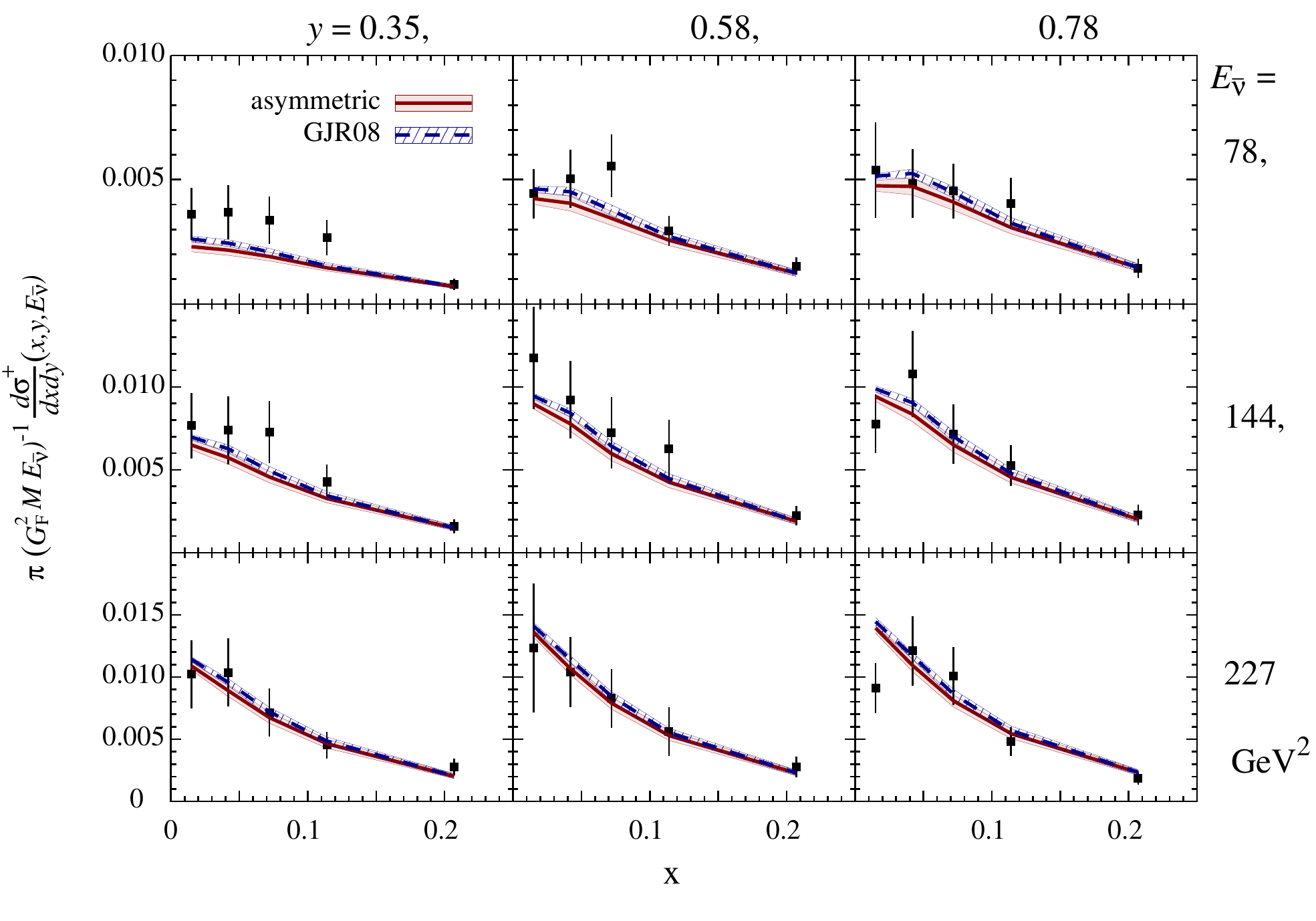}
\caption{Comparison of the NuTeV ``forward'' dimuon cross-section data \cite{Mason:2007zz} with predictions using the (strange--symmetric) GJR08 distributions as well as our newly determined ``asymmetric'' results.\label{data}}
\end{figure}

The results obtained with the GJR08 distributions are shown in Fig.\ref{data}, where a good agreement is observed; the $\chi^2$ obtained is 65 for 90 data points\footnote{Since the NuTeV collaboration has inflated the errors as a way of (not) taking into account the point to point correlations, this level of agreement briefly corresponds to 1\,$\sigma$ \cite{Goncharov:2001qe}.}. This agreement demonstrates the compatibility of the data with the conditions of Eq.~(\ref{nullstrange}) and shows that in the dynamical case, where the input distributions are parametrized at an optimally chosen low input scale $Q_0^2\!=\!0.5$ GeV$^2$ \cite{Gluck:2007ck}, the strange sea can be generated entirely radiatively starting from:
\begin{equation}
s^+(x,Q_0^2)\equiv s(x,Q_0^2)+\bar{s}(x,Q_0^2)=0 
\label{splusnull} 
\end{equation}

In addition, the small differences between neutrino and antineutrino data are well known to induce a small asymmetry in the strange sea \cite{Barone:1999yv, Olness:2003wz, Mason:2007zz, Lai:2007dq, Alekhin:2008mb, Martin:2009iq, Ball:2009mk}. In order to evaluate this asymmetry within our framework, we parametrize a new input distribution:
\begin{equation}
s^-(x,Q_0^2)\equiv s(x,Q_0^2)-\bar{s}(x,Q_0^2)= N x^a (1-x)^b (1-\tfrac{x}{x_0})
\label{sminus} 
\end{equation}
where $Q_0^2\!=\!0.5$ GeV$^2$ is (fixed to) the input scale of our GJR08 NLO fit \cite{Gluck:2007ck} and the function is constrained by the quark--number sum rule $\int_0^1dx\; s^-\!(x,Q_0^2)=0$. Eqs.~(\ref{splusnull})--(\ref{sminus}) imply that the strange input distributions, $s(x,Q_0^2)$ and $\bar{s}(x,Q_0^2)$, will in turn be negative (positive) at the input scale, by construction. This is not a problem as long as at perturbative scales, say for $Q^2\!>\!1$ GeV$^2$, both strange distributions $s(x,Q^2)$ and $\bar{s}(x,Q^2)$ become manifestly positive due to the QCD evolution, as is the case.

We have tried several (more elaborated) parametrizations without finding any improvement. As a matter of fact, even this ``minimal'' parametrization showing the required properties (vanishing at small and large $x$ and crossing the $x$--axis at least once) is too flexible for the restricted $x$--range of the data that we are using ($0.02\lesssim x \lesssim 0.3$). In practice there is a high correlation between $a$ and $b$, so that both parameters should not be kept free simultaneously, in particular for the error estimation. We have chosen to fix $b\equiv 25$, which is comparable to the value obtained for the analogous parameter of the light--quark sea--asymmetry input distribution $b_{\bar{d}-\bar{u}}\!\simeq\!17$ \cite{Gluck:2007ck}, but a bit larger as would be expected for the (heavier) strange quark; several values in the range 20 to 30 give similar results. This leaves us with 2 free parameters for the strangeness asymmetry; the crossing point of the input asymmetry $x_0$ is determined through the no--net--strangeness condition.  The optimal values of these parameters have been determined following \cite{Gluck:2007ck}; we get $N\!=\!-0.017\pm0.016$ and $a\!=\!0.22\pm0.09$. 

Since the rest of the data used in our global fit \cite{Gluck:2007ck} are essentially insensitive to an asymmetry in the strange sea\footnote{Actually, there are some marginal dependences, e.g. in $F^p_3$ or in the Drell--Yan cross--sections, which are completely negligible, for example, their effect in $\chi^2$ is well within the rounding error of the values quoted in \cite{Gluck:2007ck}.} and, moreover, we continue to generate the strange sea entirely radiatively through Eq.~(\ref{splusnull}), the parameters quoted in \cite{Gluck:2007ck} remain unchanged and our present result for the asymmetry can be used together with the GJR08 distributions. The evaluation of uncertainties is performed as in \cite{Gluck:2007ck}, however, since the free parameters introduced here are completely uncorrelated with the ones determined in \cite{Gluck:2007ck}, and there is no compatibility problems with the data, we do not find necessary to increase the tolerance parameter and use $T\!=\!1$ for the 1\,$\sigma$ uncertainties of $s^-(x,Q^2)$.

The newly obtained ``asymmetric'' results are also shown in Fig.~\ref{data}, where it can be seen that the (anti)neutrino data prefer (smaller)larger values, i.e. the data favor an increase of the $s$ distribution and a decrease of the $\bar{s}$, in other words, a \emph{positive} asymmetry in the relevant $0.01\lesssim x \lesssim 0.1$ region; for $x$ values larger than about $0.1$ no significant changes are appreciated. After introducing the asymmetry, the $\chi^2$ value improves to 60 for 90 data points, although the predictions from the strange--symmetric GJR08 distributions are rather similar and the differences lie within the 1~$\sigma$ bands. 

\begin{figure}[t]
\centering
\includegraphics[width=0.92\textwidth]{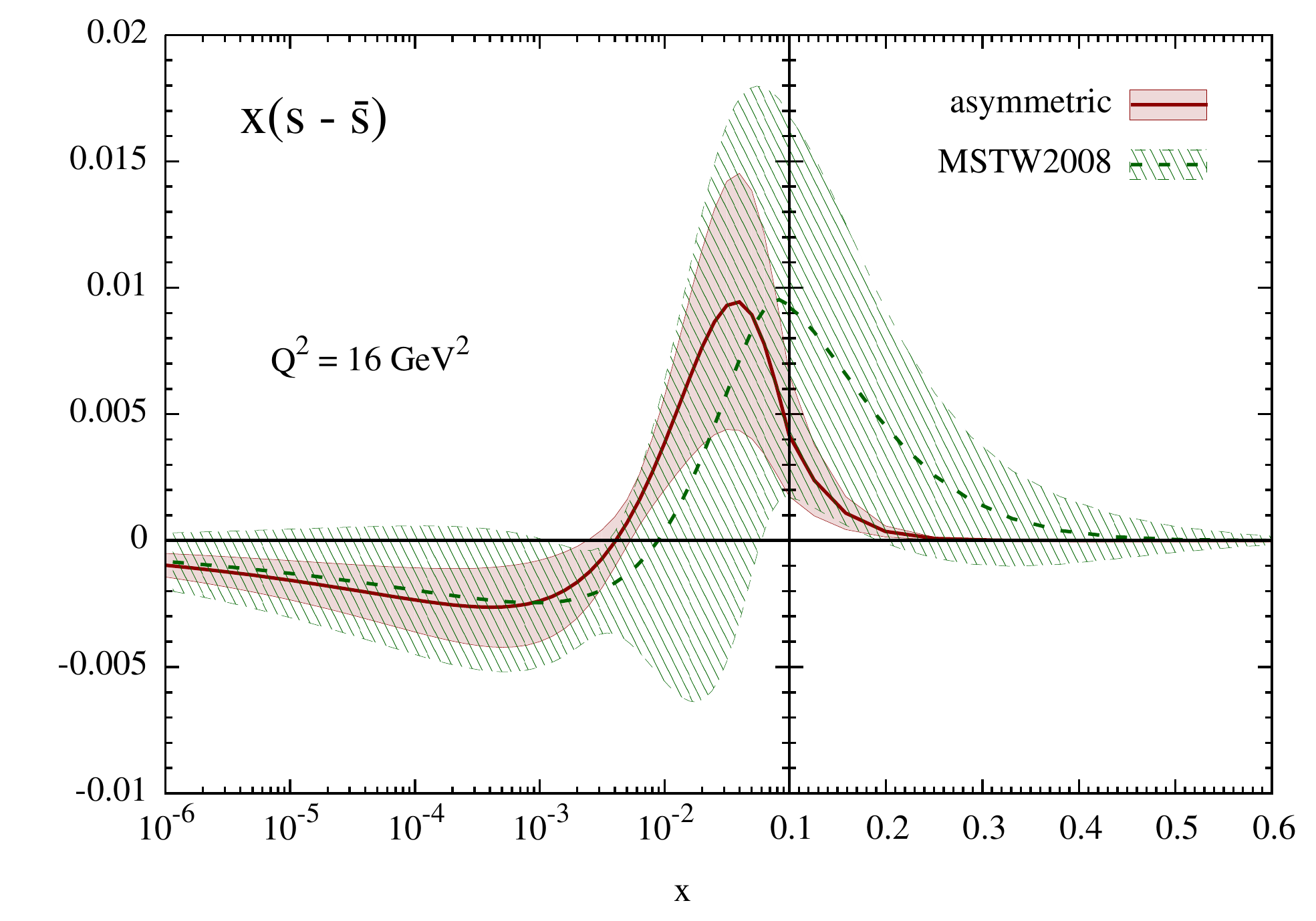}
\caption{Out result for the strangeness asymmetry in the nucleon at $Q^2\!=\!16$ GeV$^2$ appropriate for the NuTeV experiment (cf. Fig.~3 of \cite{Mason:2007zz}). The results of MSTW2008 \cite{Martin:2009iq} are also shown for comparison.\label{sasymmetry}}
\end{figure}
Our result for the strangeness asymmetry in the nucleon are shown in Fig.~\ref{sasymmetry} at $Q^2\!=\!16$ GeV$^2$ appropriate for the NuTeV experiment, and can be directly compared with Fig.~3 of \cite{Mason:2007zz}. Although due to the large errors both results are in general agreement, the peak of our asymmetry is lower and placed at a slightly smaller value of $x$. The results of MSTW2008 \cite{Martin:2009iq} are also shown in Fig.~\ref{sasymmetry} for comparison. They are rather similar in size to ours, despite the fact that in \cite{Martin:2009iq} older (and less precise) data from CCFR \cite{Goncharov:2001qe} have also been included and this tends to reduce the asymmetry \cite{Alekhin:2008mb}. Note, however, that our asymmetry is much more suppressed for large $x\!\gtrsim\!0.2$, where the data are in excellent agreement with our (strange--symmetric) GJR08 distributions \cite{Gluck:2007ck}, as can be seen in Fig.~\ref{data}. 

The changes in the strange--asymmetric distributions as compared with the original GJR08 are rather small, e.g. at $Q^2\!=\!100$ GeV$^2$ they reach at most $5\%$ in the relevant $10^{-3} \!<\! x\! <\! 0.3$ region, and are comparable with the uncertainties in the distributions, which are of a few percent as well. This being the case, the original strange--symmetric GJR08 distributions should suffice for most applications, moreover since most observables depend essentially only on $s^+(x,Q^2)$. Furthermore, since we continue to generate the strange distributions radiatively starting from Eq.~(\ref{splusnull}), the increase in the uncertainties encountered in common ``standard'' fits, where $s^+(x,Q_0^2)$ has to be fitted, is avoided in the more constrained dynamical case, which uncertainties should be very similar to the ones of GJR08 in most cases. 

The strangeness asymmetry is however relevant for applications especially sensitive to the strange content of the nucleon, as has been shown, for instance, in relation with the so--called the ``NuTeV anomaly'' (see, e.g. \cite{Gluck:2005xh} and references therein). As indication of the size and sign of the asymmetry it has become conventional to use the value of its second moment at the reference scale $Q^2\!=\!20$ GeV$^2$, we obtain:
\begin{equation}
S^- \equiv \int_0^1 dx\; x(s-\bar{s})=0.0008 \pm 0.0005
\end{equation}
which is of the right sign and size as to explain the ``anomaly'' and furthermore has, as expected, a relatively small error due to the dynamical assumptions. Previous determinations \cite{Mason:2007zz, Alekhin:2008mb, Martin:2009iq} generally yield a larger value of about 0.0010 to 0.0020 and a typical uncertainty of about $100\%$ or even larger \cite{Ball:2009mk}.

In conclusion, although in our global QCD fits \cite{Gluck:2007ck, JimenezDelgado:2008hf} no data with especial sensitivity to the strange content of the nucleon have been included, our determination of strange parton distributions, in particular Eq.~(\ref{splusnull}), is compatible with particularly sensitive data, e.g. those in \cite{Mason:2007zz}. Furthermore, these data induce an asymmetry in the strange sea which has been evaluated within our dynamical framework and found, in agreement with previous results, to be rather small and positive. This being the case, our original strange--symmetric distributions should suffice for most applications. The strangeness asymmetry may, however, be relevant for some especially sensitive applications; for these cases our results are available on request\footnote{pjimenez@physik.uzh.ch}.

\section*{Acknowledgements}
We would like to thanks E.~Reya  and T.~Gehrmann for carefully reading the manuscript and useful discussions, S.~Kretzer for code, and S.~Alekhin and D.~de Florian for providing the acceptance corrections and the nuclear corrections respectively. This work has been supported in part by the ``Bundesministerium f\"ur Bildung und Forschung'', Berlin; and by the Swiss National Science Foundation (SNF) under contract 200020-126691.

\end{document}